\documentclass{amsart}       

\usepackage{hyperref}
\usepackage{graphicx}
\usepackage{fixltx2e}                    
\usepackage[utf8]{inputenc}            
\usepackage{amssymb,latexsym, dsfont, amsfonts, amsbsy, amsmath, mathrsfs, dsfont}            
\usepackage{mathtools}                   
\usepackage{caption}
\usepackage{bm}                          
\usepackage{enumerate}                   
\usepackage{verbatim}                    
\usepackage{url}                         
\usepackage{tabularx}
\usepackage{microtype}  
\usepackage[all, knot]{xy}
\xyoption{arc} 
\xyoption{web}                 
\makeatletter                            
\def\MT@register@subst@font{\MT@exp@one@n\MT@in@clist\font@name\MT@font@list
	\ifMT@inlist@\else\xdef\MT@font@list{\MT@font@list\font@name,}\fi}
\makeatother

\usepackage{color}

\usepackage{scalerel,stackengine}
\stackMath
\newcommand\what[1]{%
	\savestack{\tmpbox}{\stretchto{%
			\scaleto{%
				\scalerel*[\widthof{\ensuremath{#1}}]{\kern-.6pt\bigwedge\kern-.6pt}%
				{\rule[-\textheight/2]{1ex}{\textheight}}
			}{\textheight}%
		}{0.5ex}}%
	\stackon[1pt]{#1}{\tmpbox}%
}

\newcommand{\const}[1]{\overline{#1}}
\newcommand{\termDef}[1]{\textbf{\textit{#1}}}

\newcommand{\?}{\ensuremath{\mkern0.4\thinmuskip}}   

\let\leq=\leqslant
\let\geq=\geqslant
\let\Box=\square                            

\newcommand{\N}{\mathbb{N}}

\newcommand{\depth}{\mathbf{\mathtt{d}}}

\newcommand{\sL}{\text{\footnotesize{\L}}}
\newcommand{\ssL}{\text{\scriptsize{\L}}}

\def\SF{\textit{SFm}}
\def\PSF{\textit{PSFm}}

\let\alg=\bm                                    
\let\aclass=\mathcal                              
\let\class=\mathbb                             
\let\mod=\mathfrak							
	
\let\alg=\mathbf

\let\epsilon=\varepsilon
\let\Lambda\varLambda
\let\Gamma\varGamma
\let\Delta\varDelta
\let\Lambda\varLambda
\let\Omega\varOmega
\let\Theta\varTheta
\let\Xi\varXi
\let\Pi\varPi
\let\Sigma\varSigma

\newcommand{\aFL}{\mathbf{FL_{ew}}}                   
\newcommand{\FL}{\mathcal{F}L_{ew}}                  

\newcommand{\lu}{\scriptsize{\L}}

\theoremstyle{plain}
\newtheorem{theorem}{Theorem}[section]
\newtheorem{lemma}[theorem]{Lemma}
\newtheorem{corollary}[theorem]{Corollary}
\newtheorem{proposition}[theorem]{Proposition}

\theoremstyle{definition}

\newtheorem{definition}[theorem]{Definition}
\theoremstyle{remark}

\begin{document}
\title{On transitive modal many-valued logics}
\author{Amanda Vidal}

\maketitle

\vspace{-0.8cm}
\begin{center}
\textit{ Institute of Computer Science, Czech Academy of Sciences \\
Pod Vodárenskou v{\v e}{\v z}\'i 271/2
182 07. Prague, Czech Republic}\\
\textsf{amanda@cs.cas.cz}
\end{center}
\vspace{0.2cm}

\thispagestyle{empty}

\begin{abstract}
	This paper is focused on the study of modal logics defined from valued Kripke frames, and particularly, on computability and expressibility questions of modal logics of transitive Kripke frames evaluated over certain residuated lattices. It is shown that a large family of those logics -including the ones arising from the standard MV and Product algebras- yields an undecidable consequence relation. Later on, the behaviour of transitive modal \L ukasiewicz logic is compared with that of its non transitive counterpart, exhibiting some particulars concerning computability and equivalence with other logics. We conclude the article by showing the undecidability of the validity and the local SAT questions over transitive models when the $\Delta$ operation is added to the logic.
%
%
%
%
\end{abstract}
\section{Introduction}

Modal logic is one of the most developed and studied non-classical logics, yielding a beautiful equilibrium between computational complexity and expressibility. Generalizations of the concepts of necessity and possibility offer a rich setting to model and study notions from many different areas, including proof-theory, temporal and epistemic concepts, work-flow in software applications, etc. On the other hand, substructural logics 
provide a formal framework to manage vague and resource sensitive information in a very general (and so, adaptable) fashion.

 \textit{Modal many-valued  logics} appear in the literature both pursuing purely theoretical development and also with the objective of 
 offering a richer framework to model complex environments that might require valued information as well as qualification operators. 
 While the first publications on the topic can be traced back to the 90s \cite{Fi92a,Fi92b} (that focus on the problem over finite Heyting algebras), it has been only in the latter years when a more systematic work has been developed. In \cite{Ha98} a brief study of the S5 modal logics over BL algebras is presented, but it is in more recent works where the modal logics over arbitrary Kripke frames (also referred to in the literature as \textit{minimal modal logics}) are studied. 
 
 Several works since have studied different aspects of these logics. Most relevant for the present paper are the works related to axiomatizability and proof-theoretic questions, addressing the minimal modal logics over finite  MTL algebras \cite{BoEsGoRo11}, \L ukasiewicz finite and infinite standard algebras \cite{HaTe13}, Product standard algebra \cite{ViEsGo16}, and Gödel standard algebra \cite{CaRo10, CaRo15}, \cite{MeOl11}.
 
 Concerning computability, in \cite{CaMe13,CaMeRo17} it is proven that the minimal (local) modal Gödel logics with both $\Box$ and $\Diamond$ modal operators are decidable (both over models with a crisp accessibility relation and with a valued one). It is also shown that the S5 extension of the previous logic with crisp accessibility (equivalent to the one-variable fragment of predicate Gödel logic) is decidable too. However, in relation to the present paper, we point out that the question whether the purely transitive extension is decidable or not is left open.
 For modal \L ukasiewiccz and Product logics, no general results on decidability have been proven, and the failure of the finite model property, as well as the difficulties to get recursive and finitary axiomatizations for them make the possible answers to this question non trivial to conjecture. 

The nearest problem addressed in the literature concerns the decidability of some Fuzzy Description Logics (FDL) (see eg. \cite{St01}, \cite{Ha05b}, \cite{BaPe11a}, \cite{CeSt13}, \cite{BoDiPe15}). These logics expand towards the valued setting the so-called Description Logics, a formalism used intensively  in AI and ontologies which can be seen as semantic variations (in some cases, also syntactic) of modal logic. In relation to fuzzy modal logic, we can see FDL as a multi-modal system over models with both weighted accessibility relations and formulas, that is not based on the complete usual logical language but that has, on the other hand, names for worlds and the possibility of referring (via constants) to each element of the algebra of evaluation. The study of decision procedures in FDL is focused in variants of the r-SAT (existence of a valuation that valuates to at least $r$) problem, and in \cite{CeEs18} we can find a translation of the known results to the context of many-valued modal logics. However, since these results are limited to the context of valued accessibility relation and multi-modal operations, it does not seem likely to exist a uniform translation of them 
to modal logics arising from classical frames with valued formulas, the topic of study in the ongoing work. Moreover, questions concerning validity and derivability in the logic remain, in most cases, open.\footnote{It is known from \cite{Ha05b} that  validity over the multi-modal \L ukasiewicz logic with fuzzy accessibility relation is decidable, and a similar result concerning the product case was presented with partial mistakes in \cite{CeEsBo10}, and corrected in unpublished notes by the authors.} A general approach to determine undecidability of consistency  over FDLs is developed in \cite{BoDiPe15}, proving in particular that the SAT problem over Product and \L ukasiewicz FDLs is undecidable as long as certain expressivity conditions are met. 
However, the approach is not suited to cope with the problems studied in this paper, since they belong to non-comparable settings. On the one hand, our main goal is that of shedding some light over decidability of the minimal logics (both as sets of theorems or as deduction systems) arising from valued models with a crisp accessibility relation. On the other hand, the methods from the previous reference are focused on the question of consistency (nor reducible to validity since the logic is many-valued) and moreover, strongly related to the language of FDLs (with incorporates eg. constants for the elements of the models) and the possibility of assigning degrees to the accessibility relations, none of which can be done in our context.

Along this paper, we focus on the study of the decidability of the local consequence relation on modal logics over models with crisp accessibility relation valued on certain classes of $FL_{ew}$-algebras, that comprehend the well-known cases of 
the \L uaskewicz standard algebra, the class of finite MV chains, the standard Product algebra and the one-generated product algebra. The main contribution of the paper is that the consequence over transitive models of the above kind are undecidable, also if we restrict the logic to the one arising from only the finite models in the class. Remarkably enough, transitive models are one of the most common kind of relational models naturally  appearing in CS and other fields (from accessibility models of the real world to dynamic-logic style software formalizations, preferences and other epistemic notions modelling, etc). Thus, the undecidability of these logics points to the problems that might arise with their use for applications in an unrestricted way, as well as opens to consideration the study of weaker logics with better computational behaviour. 

A second main contribution of this paper is an study of some particularities of the modal logics defined extending propositional \L ukasiewicz logics. First, arising as a consequence of some results from \cite{Ha05b} and \cite{Ca18}, we show the decidability of the local modal \L ukasiewicz logic (as consequence relation), which interestingly  provides us with an example of a decidable modal logic whose transitive expansion is undecidable (a phenomena of which, to the best of our knowledge, there were not known examples up to now). On the other hand, we also observe that, while the minimum (local) modal logic over the standard MV algebra, and that over all finite MV algebras coincide, this is not the case for the respective transitive logics.

The paper is structured as follows: In section \ref{sec:prel} we introduce all the definitions that will be used throughout the paper, aiming to be as self-contained as possible. Section \ref{sec:undec} focuses on the undecidability result stated above, and details the reduction of the logical consequence over transitive models to the Post Correspondence Problem. Section \ref{sec:lukDec} shows the decidability of the local modal \L ukasewicz logic, and provides a separating example for transitive modal logic over the standard MV algebra and the one over all finite MV chains. Lastly, in Section \ref{sec:delta} we observe how the previous logics expanded with the Monteiro-Baaz $\Delta$ operation turn to have not only  undecidable consequence relation, but also undecidable validity and SAT.

\section{Preliminaries}\label{sec:prel}

Modal many-valued logics arise from Kripke structures evaluated over certain algebras, putting together relational and algebraic semantics in a fashion adapted to model different reasoning notions. Along the present work, the algebraic  basis of these semantics will be the one of $FL_{ew}$-algebras, the corresponding algebraic semantics of the Full Lambek Calculus with exchange and weakening \cite{GaJiKoOn07},\cite{DoSH93}. This will offer a very general approach to the problem while relying in well-known algebraic structures. Along this section, we will  formally introduce the previous algebras and the basic definitions necessary for the further development of the paper.

\begin{definition}
	A \termDef{$\aFL$-algebra} is a structure $\alg{A} = \langle A;  \wedge, \vee, \cdot, \rightarrow, 0, 1\rangle$ such that
	\begin{itemize}
		\item $\langle A; \wedge, \vee, 0,1\rangle$ is a bounded lattice;
		\item $\langle A; \cdot, 1\rangle$   is a commutative monoid;
		\item $\alg{A}$ satisfies $a \cdot b \leq c$  if and only if $a \leq b \rightarrow c$  for any $a,b,c\in A$. 
	\end{itemize}	
\end{definition}
We will usually write $ab$ instead of $a \cdot b$, and abbreviate $\overbrace{x \cdot x \cdots x}^{n}$ by $x^n$. Moreover, as it is usual, we will define $\neg a$ to stand for $a \rightarrow 0$. 
In the setting of the previous definition, we will denote by $\alg{Fm_p}$ the algebra of formulas built over a countable set of variables $\mathcal{V}$ using the language corresponding to the above class of algebras (i.e., $\langle \wedge/2, \vee/2, \cdot/2, \rightarrow/2, 0/0, 1/0\rangle$). As usual, we let \[(x \leftrightarrow y) \coloneqq x \rightarrow y) \cdot (y \rightarrow x) \quad \text{ and }\quad  \neg x \coloneqq x \rightarrow 0.\]

Let us introduce some well-known examples $\aFL$-algebras over the universe $[0,1]$ (in fact, also BL algebras, i.e, further satisfying prelinearity -MTL- and divisibility \cite{Ha98},\cite{EsGo01}). In the algebras below, $\wedge$ and $\vee$ stand for usual lattice conjunction (min) and disjunction (max) in $[0,1]$, and all standard algebras have as universe the real unit interval $[0,1]$. Then 
\begin{itemize}
	\item $[0,1]_G$, the \termDef{standard Gödel algebra}, further lets  
	\[a \cdot b \coloneqq a \wedge b \qquad \text{and} \qquad a \rightarrow b \coloneqq \begin{cases} 1 &\hbox{ if } a \leq b\\ b &\hbox{ otherwise} \end{cases}\]
	
	\item $[0,1]_{\sL}$, the \termDef{standard MV algebra}, further lets 	\[a \cdot b \coloneqq \max\{0, a + b -1\} \qquad \text{and} \qquad a \rightarrow b \coloneqq \min\{1, 1 - a + b\}\]
	\item $MV_n$, the \termDef{finite MV algebra} of $n+1$ elements, is the subalgebra of $[0,1]_{\sL}$ with respect to the subuniverse $\{0, \frac{1}{n}, \ldots, \frac{n}{n}\}$;
	
	
	\item $[0,1]_\Pi$, the \termDef{standard Product algebra}, further lets 
	\[ a \cdot b \coloneqq a \times b \qquad \text{and} \qquad  a \rightarrow b \coloneqq \begin{cases} 1 &\hbox{ if } a \leq b\\ b/a &\hbox{ otherwise} \end{cases}\] for $\times$ the usual product between real numbers;
	\item $\alg{A} \preceq_1 [0,1]_\Pi$, \termDef{one-generated product algebras} (all are isomorphic), is any subalgebra of $[0,1]_\Pi$ with universe $\{0,1\} \cup \bigcup_{i \in \omega} a^i$ for some $a \in (0,1)$.
\end{itemize}

Le us also point out some  particular characteristics of some $\aFL$-algebras that will be of use later.
\begin{definition} Let $\alg{A}$ be a $\aFL$-algebra.
	\begin{itemize}
		\item $\alg{A}$ is \termDef{$\boldsymbol{n}$-contractive} whenever $a^{n+1} = a^n$ for all $a \in A$.
		\item $\alg{A}$ is \termDef{weakly-archimedean} if for any two elements $a,b \in A$, if $a \leq b^n$ for all $n \in \omega$ then $ab = a$.
	\end{itemize}
\end{definition}

Observe that if $\alg{A}$ is n-contractive, the element $a^n$ is idempotent for any $a \in A$. Simple examples of these algebras comprehend Heyting and G\"odel algebras, and $MV_n$ algebras. On the other hand, the (infinite) standard MV-algebra, the standard product algebra and any one-generated subalgebra of the latter one are not $n$-contractive for any $n$.

For what concerns weak-archimedeanicity, observe that if the element $\inf{b^n}$ exists in a weakly-archimedean algebra, then  it is an idempotent element. Examples of weakly-archimedean algebras are the standard MV-algebra, the standard product algebra, as well as the algebras belonging to the generalised quasi-varieties generated by  them. In particular, any (non-trivial) one-generated subalgebra of the standard product algebra is weakly-archimedean .



For what concerns this work, it is interesting to recall that the logic $\aclass{FL}_{ew}$, the Full Lambek Calulus with exchange and weakening, is complete with respect to the class of logical matrices $\{\langle \alg{A}, \{1\} \colon \alg{A} \in \FL\}$. That is to say, for any $\Gamma, \varphi \subset_{\omega} Fm_p$,\footnote{The notation $\subset_{\omega}$ denotes, as usual, a finite subset.} 
\[\Gamma \vdash_{\aclass{FL}_{ew}} \varphi \text{ iff }\forall \alg{A} \in \FL, \forall h \in Hom(\alg{Fm_p}, \alg{A}),\ h([\Gamma]) \subseteq \{1\} \text{ implies } h(\varphi) = 1\]

The  algebra of modal formulas $\alg{Fm}$ will be built in the same way as $\alg{Fm_p}$, but by expanding the language of $\aFL$-algebras with two unary operators $\Box$ and $\Diamond$. While it is clear how to extend a propositional evaluation from $\mathcal{V}$ into an $FL_{ew}$-algebra to $Fm_p$, the semantic definition of the modal operators is defined from the relational structures in the following way.

\begin{definition}\label{def:AKmodel}
	Let $\alg{A}$ be a $\aFL$-algebra. An \termDef{$\alg{A}$-Kripke model} is a structure $\mod{M} = \langle W, R, e\rangle$ such that
	\begin{itemize}
		\item $\langle W, R\rangle$ is a Kripke frame.  That is to say, $W$ is a non-empty set of so-called worlds and $R$ is a binary relation over $W$, called \textit{accessibility relation};
		\item $e\colon \mathcal{V} \times W \rightarrow A$. $e$ is extended to $Fm_p$ in such a way that (world-wise) it is a homomorphism into $\alg{A}$, and to $Fm$ by further letting
		\[e(v, \Box \varphi) \coloneqq \bigwedge\limits_{\langle v,w\rangle \in R} e(w, \varphi) \qquad \text{ and } \qquad e(v, \Diamond \varphi) \coloneqq \bigvee\limits_{\langle v,w \rangle \in R} e(w, \varphi)\]
		whenever that infima/suprema exist, and undefined otherwise.
	\end{itemize}
\end{definition}

To lighten the notation, we will usually write $Rvw$, and say in this case that $w$ is a  successor of $v$, to denote $\langle v,w \rangle \in R$.

\begin{definition}
	\begin{enumerate}
		\item 	A model is \termDef{safe} whenever the values of $e(v, \Box \varphi)$ and $e(v, \Diamond \varphi)$ are defined for any formula at any world. We will denote by \termDef{$\aFL$-Kripke models} to the class of all $\alg{A}$-Kripke models for any $\aFL$-algebra $\alg{A}$. 
		\item A safe model is \termDef{witnessed} whenever for any modal formula $\texttt{M}\varphi$ and each world $v \in W$, there is $w_{\texttt{M}\varphi} \in W$ such that $Rvw_{\texttt{M}\varphi}$ and $e(v, \texttt{M}\varphi) = e(w_{\texttt{M}\varphi}, \varphi)$.
	\end{enumerate}
\end{definition}
For what concerns notation, given a class of models $\class{C}$, we denote by $\omega\class{C}$ the \textbf{finite models} in $\class{C}$ (observe these are always safe and witnessed). On the other hand, for a class of algebras $\aclass{C}$ (or a single algebra $\alg{A}$) we write $\class{K}_{\aclass{C}}$  (correspondingly $\class{K}_{\alg{A}}$) to denote the  class of safe Kripke models over the algebras in the class (or over the single algebra specified). Finally, in order to lighten the reading, we will let $\class{K}_{\sL}$, $\class{K}_{\omega\sL}$  and $\class{K}_{\Pi}$ to denote respectively $\class{K}_{[0,1]_{\tiny{\L}}}$, $\class{K}_{\{MV_n\colon n \in \omega\}}$ and $\class{K}_{[0,1]_\Pi}$.

As it happens for classical models, we can also consider some condition only over the kind of accessibility relation and study the logic arising from the corresponding classes of models. Along this work, we are focused in the restriction to transitive accessibility relations, i.e., those models such that for any $u,v,w \in W$, if $Ruv$ and $Rvw$ then $Ruw$. As usual, for an arbitrary class of models  $\class{C}$, we will denote the transitive models in it by $4\class{C}$. Observe, however, this is only a naming convention, since we are not assuming in any case that the transitive logic corresponds to an extension of the minimal one by the $4$ axiom(s) schemata.

Towards the definition of modal logics over $\aFL$-algebras relying in the notion of $\aFL$-Kripke models, it is natural to preserve the notion of world-wise truth being $\{1\}$ (in  order to obtain, if restricted to world-wise, the propositional $\aclass{FL}_{ew}$ logic). With this in mind, for any $\alg{A}$-Kripke model $\mod{M}$  and $v \in W$  we say that $\mod{M}$ \termDef{satisfies a formula} $\varphi$ \termDef{in $v$}, and write $\mod{M},v \models \varphi$, whenever  $e(v, \varphi) = 1$. Similarly, we simply say that $\mod{M}$ \termDef{satisfies a formula} $\varphi$, and write $\mod{M} \models \varphi$ whenever for all $v \in W$ $\mod{M},v \models \varphi$.

Over the previous notion of satisfiability, two different consequence relations can be defined, a local and a global one. Along the present work, we will focus on the preservation of truth \textit{locally}.

%
\begin{definition} Let $\Gamma, \varphi \subseteq_{\omega} Fm$, and $\class{C}$ be a class of $\aFL$-Kripke models. Then we say that
%
		\termDef{$\varphi$ follows from $\Gamma$ locally in  $\class{C}$}, and we write $\Gamma \vdash_{\class{C}} \varphi$, whenever for any $\mod{M} \in \class{C}$ and any $v \in W$, \[\mod{M}, v \models \Gamma \text{ implies }\mod{M}, v \models \varphi;\]
	When $\class{C}$ is clear from the context, we will simply write $\vdash$. Moreover, for a model $\mod{M}$ and a world $v \in W$, we will write $\Gamma \not \vdash_{\langle \mod{M}, v \rangle} \varphi$ to denote that $e(v, \Gamma) \subseteq \{1\}$ and $e(v, \varphi) < 1$. 
\end{definition}

Observe the necessity rule $\varphi \vdash \Box \varphi$ is only valid in the above deductive system for theorems of the logic, as it happens in the classical local modal logic.

The following basic notions concerning manipulation of Kripke models will be of use later on.

\begin{definition}\label{lemma:depth}
	Given a Kripke model $\mod{M}$ and $w \in W$, we let the \termDef{depth of $w$} be given by
	\[\depth(w)  \coloneqq sup\{k \in \N:\?  \exists w_0, \ldots, w_k \text{ with } w_0 = w \text{ and } Rw_i w_{i+1} \text{ for all } 0 \leq i < k \}.\]
\end{definition}

Observe that if there exists some cycle in the model, all worlds involved in it have infinite depth.

\begin{definition}
	We let the \termDef{propositional subformulas of $\varphi$} be the set defined by
	\begin{eqnarray*}
		\PSF(p) & \coloneqq & \{p\}, \text{ for $p$ propositional variable or constant}\\
		\PSF(\texttt{M}\varphi) & \coloneqq & \{\texttt{M}\varphi\} \hbox{ for } \texttt{M} \in \{\Box, \Diamond\}\\
		\PSF(\varphi_1 \divideontimes \varphi_2) & \coloneqq & \SF(\varphi_1) \cup \SF(\varphi_2) \cup \{ \varphi_1 \divideontimes \varphi_2\} \hbox{ for } \divideontimes \in \{\wedge, \vee, \rightarrow \}
	\end{eqnarray*}
	For $\Gamma$ a set of formulas we let $\PSF(\Gamma) \coloneqq \bigcup_{\gamma \in \Gamma} \PSF(\gamma).$
\end{definition}

Let us finish the preliminaries by stating a well-known undecidable problem, that will be used in the next sections to show undecidability of some of the modal logics introduced above. Recall that given two numbers  $\mathtt{x}, \mathtt{y}$ in base $s \in \omega$, their concatenation $\mathtt{xy}$ is given by $\mathtt{w_1}\cdot s^{\parallel\mathtt{y}\parallel} + \mathtt{y}$ (for $\cdot, +$ the usual real product and sum), where $\parallel\mathtt{y}\parallel$ is the number of digits of $\mathtt{y}$ in base $s$.
\begin{definition}[\textbf{Post Correspondence Problem (PCP)}]\label{def:PCP}
	An instance $P$ of the PCP consists on a list $\langle {\mathtt{v_1}}, \mathtt{w_1} \rangle \dots \langle  \mathtt{v_n}, \mathtt{w_n}\rangle$ of pairs of numbers without repetitions\footnote{That is, for each $1 \leq i \neq j \leq n$ either $\mathtt{v_i} \neq \mathtt{v_j}$ or $\mathtt{w_i} \neq \mathtt{w_j}$.} in some base $s \geq 2$.
	A solution for $P$ is a sequence of indices $i_1, \dots, i_k$ with  $1 \leq i_j \leq n$ such that \[\mathtt{v_{i_1}v_{i_2}  \ldots v_{i_k}}  = \mathtt{w_{i_1} w_{i_2} \ldots w_{i_k}} . \]
\end{definition}

Finding a solution for PCP-instances yields an undecidable procedure \cite{Post46}.

\section{Undecidability of transtive local deduction}\label{sec:undec}

Along the following sections, unless stated otherwise, we let $\aclass{A}$ to be a class of weakly-archimedean linearly ordered $\aFL$ algebras
such that for any $n \in \omega$ there is some $\alg{A}_n \in \aclass{A}$ such that
$\alg{A}_n$ is non n-contractive\footnote{We conjecture that the same results hold if we remove the linearity condition. However, due to the lack of existing or natural examples from this more general framework, and the drawback that the undecidability proof gets much more cumbersome, have together led the author to avoid formulating the result in that more general fashion.}. That is to say, there is some $a \in A_n$ such that 
$a^{n+1} < a^n.$

Examples of classes of algebras like the above one are 	$\{[0,1]_{\lu}\}$ $\{MV_n\colon n \in \omega\}$, $\{[0,1]_{\Pi}\}$ and $\{\alg{A}\} $ for $\alg{A}   \preceq_1[0,1]_\Pi$. 
Natural examples of classes of algebras that do not satisfy the above conditions are $\{[0,1]_G\}$ (and the variety generated by it) and the varieties of MV and product algebras. 

By relying on the properties specified above for the class of algebras $\aclass{A}$,  we can prove the following result.
\begin{theorem}\label{th:undec}
	The problem of determining whether $\varphi$ follows locally from $\Gamma$ in $4\class{K}_{\aclass{A}}$ is undecidable. Moreover, also the problem of determining whether $\varphi$ follows locally from $\Gamma$ in $\omega4\class{ K}_{\aclass{A}}$ is undecidable. 
	More in particular, the three-variable fragments of both previous deductive systems are undecidable.
\end{theorem}

Its proof follows as a simple consequence of Proposition \ref{prop:PsatLocal}, which we now proceed to formulate and prove.
In order to do so, given an arbitrary instance $P = \{\langle \mathtt{v}_1, \mathtt{w}_1 \rangle, \ldots, \langle \mathtt{v}_m, \mathtt{w}_m \rangle\}$ 
of the Post correspondence problem, let us define a set of formulas $\Gamma_P \cup \varphi_P$. We let $\Gamma_P$ be
the union of the following formulas with variables $\mathcal{V} = \{y, v, w\}$:

\begin{enumerate}
	\item $\Box y \leftrightarrow \Diamond y$;

	\item $\Box \bigvee\limits_{1 \leq i \leq m} (v \leftrightarrow (\Box v)^{s^{\parallel \mathtt{v_i}\parallel}}  y^{\mathtt{v_i}}) \land  (w \leftrightarrow (\Box w)^{s^{\parallel \mathtt{w_i}\parallel}}  y^{\mathtt{w_i}})$;
	
	
	\item $\Box(\Box (v    w) \to (\Box v    \Box w))$
\end{enumerate}

Finally, let \[\varphi_P = (v \leftrightarrow w) \rightarrow (y \vee (v  w \to v   w   y))\]

%
%
%
%
%
%


Let us prove some technical lemmas concerning Kripke models with a world in which $\Gamma_P$ holds, but not $\varphi_P$.

First, we can easily see how variable $y$ is forcing certain conditions on the underlying structure of those models.
$\Gamma_P$ suffices to prove a completeness with respect to models where the variable $y$ takes the same value everywhere, except possibly in the root world (whose value is irrelevant for the proof).
\begin{lemma}\label{lemma:valueylocal}
Let $\mod{M} \in 4\class{K}_{\aclass{A}}$ be a transitive $\alg{A}$-Kripke model and $u \in W$ be such that $\Gamma_P \not \vdash_{\langle \mod{M},u\rangle} \Box \varphi_P$. 
Then there is $\alpha_y \in A$ such that for all $t_1, t_2 \in W$ with $Rut_1$ and $Rut_2$
\[e(t_1, y) = e(t_2, y) = \alpha_y.\]
\end{lemma}
\begin{proof}
	Assume $Rut_1$ and $Rut_2$, and towards a contradiction let $e(t_1, y) < e(t_2, y)$. Then, by definition, $e(u, y)  \leq e(t_1, y) < e(t_2, y) \leq e(u, \Diamond y)$, contradicting that $e(u, (1)) = 1$. 

\end{proof}
Since the model is transitive, this allows us to affirm that if $\Gamma_P \not \vdash_{4\class{K}_{\aclass{A}}} \Box \varphi$, then it happens in a tree $\mod{M}$ with root $u$, and so that there is $\alpha_y\in A$ such that for all world $t \in W\setminus \{u\}$, \[e(t,y) = \alpha_y.\]
We will resort to this fact below without further notice.

The way we chose both $\Gamma_P$ and $\varphi_P$ are also determining that the model (as in the above paragraph) is of finite depth. Contrary to what happens in the minimal modal logics, where the local deduction is naturally complete with respect to models of finite depth (indeed, bounded by the maximum modal depth of the formulas involved in the derivation), observe this is not the case in general for  transitive logics.
\begin{lemma}\label{lemma:finiteDepthLocal}
		Let $\mod{M} \in 4\class{K}_{\aclass{A}}$ and $u \in W$ be such that $\Gamma_P \not \vdash_{\langle \mod{M}, u \rangle} \Box \varphi_P$. 
		Then there is some $z \in W$ such that $Ruz$, $e(z, \varphi_P) < 1$ and $z$ has finite depth.
\end{lemma}
\begin{proof}
	The existence of $z \in W$ such that $Ruz$ and $e(z, \varphi_P) < 1$ follows by definition, since $e(u, \Box \varphi_P) < 1$. To prove that $z$ has finite depth, we can rely  in the formula $(2)$ from $\Gamma_P$, the previous lemma and the formula in the right side of $\varphi_P$ and prove by transfinite induction on the depth of the world that for any $t \in W$ such that $Rut$ and any $n \in \omega$, 
	\begin{equation}\label{eq:lemmafinitedepthlocal}
	\text{ If } \depth(t) \geq n \text{ then } e(t, v) \leq \alpha_y^n
	\end{equation}

	\begin{itemize}
		\item for $\depth(t) = 0$ is trivial since $\alpha_y^0 = 1$ by definition.
		\item For $\depth(t) = n+1$ there is $r \in W$ with $Rtr$ and $\depth(r) = n$. Then, for some $1 \leq i \leq m$, 
		\[e(t, v) = e(t, \Box v)^{s^{\parallel \mathtt{v_i} \parallel}}   \alpha_y^{\mathtt{v_i}} \leq e(r,v)^{s^{\parallel \mathtt{v_i} \parallel}}   \alpha_y^{\mathtt{v_i}} \qquad \text{ from (2) in }\Gamma_P.\]
		By  Induction  Hypothesis, and since $P$ does not have empty words, the previous is  less or equal than  $\alpha_y^{n}   \alpha_y$, and so, $e(t,v) \leq \alpha_y^{n+1}$, proving the step.
		\item Assume $\depth(t) = \omega$. Then, for any $n \in \omega$, there is some $r_n \in W$ with $Rtr_n$ and $\depth(r_n) \geq n$. 
		As before, 	\[e(t, v) = e(t, \Box v)^{s^{\parallel \mathtt{v_i} \parallel}}   \alpha_y^{\mathtt{v_i}}\]
		and so, \[e(t,v) \leq e(r_n, v) \qquad \text{ for all }n \in \omega.\]
		By induction hypothesis, $e(r_n, v) \leq \alpha_y^n$, and so, $e(t,v) \leq \alpha_y^n$ for all $n \in \omega$.
	\end{itemize}
Now, assume towards a contradiction that $\depth(z)$ were to be infinite. From condition (\ref{eq:lemmafinitedepthlocal}) it would hold that $e(z,v) \leq \alpha_y^n$ for all $n \in \omega$. Since the algebras in $\aclass{A}$ were required to be weakly-archimedean, we know this implies that $e(z, v)   e(z, y) = e(z, v)$. However, since $e(z,\varphi_P) <1$, in particular necessarily $e(z, v \rightarrow v   y) < 1$, contradicting the assumption and proving the lemma.

%
%
%
\end{proof}

At this point, we have proven completeness with respect to to trees of finite depth (by simply taking a model given by the root, the world identified in the previous Lemma, and all the successors of it). We can now turn our attention to the behaviour of variables $v,w$ y that model.
	\begin{lemma}\label{lemma:valuevpower}
	Let $\mod{M} \in 4\class{K}_{\class{A}}$ be a tree of finite depth with root $u$ such that $\Gamma_P \not \vdash_{\langle \mod{M}, u \rangle} \Box \varphi_P$, and $z$ be as in the previous lemma. 
	Then, for each $r \in W$ with $Rzr$ or $r=z$, there are $a_r, b_r \in \omega$ for which
	\[e(r,v) = \alpha_y^{a_r} \qquad \text{ and } \qquad e(r,w) = \alpha_y^{b_r}. \]
	Moreover, if $Rtr$ then $a_r < a_t$ and $b_r < b_t$.
\end{lemma}
\begin{proof}
We can prove it by induction in the depth of $r$. We do the case for $v$, the other one is analogous:
\begin{itemize}
	\item if $\depth(r) = 0$, then from $(2)$  in $\Gamma_P$ it holds there is some $1\leq i \leq m$ for which $e(r, v) = e(r, \Box v)^{s^{\parallel \mathtt{v_i} \parallel}}   \alpha_y^{\mathtt{v_i}}$, thus $e(r, v) = 1   \alpha_y^{\mathtt{v_i}}$.
	\item For $\depth(r) = n+1$, again by $(2)$ and applying I.H, there is some $1 \leq i \leq m$ for which
	\[e(r, v) = (\bigwedge_{Rrt} e(t, v))^{s^{\parallel \mathtt{v_i} \parallel}}   \alpha_y^{\mathtt{v_i}} = (\bigwedge_{Rrt} \alpha_y^{a_t})^{s^{\parallel \mathtt{v_i} \parallel}}   \alpha_y^{\mathtt{v_i}}\]
	(with $a_t \in \omega$).

	Observe that $|\{a_t\colon Rrt\}| = \omega$ would imply that $e(r,v) \leq \alpha_y^n$, and thus $e(z, v) \leq \alpha_y^n$, for all $n \in \omega$. Then, by the same reasoning from the previous lemma, we would get a contradiction with $e(z, \varphi_P) < 1$. This implies that necessarily $|\{a_t\colon Rrt\}|$ is a finite set, and so it has a maximum element $a$.  Thus, 
	\[e(r,v) = (\alpha_y^{a})^{s^{\parallel \mathtt{v_i} \parallel}}   \alpha_y^{\mathtt{v_i}}\]
	proving the first part of the lemma.
\end{itemize}
The last claim is a simple conclusion of the above relying in the fact that $e(z, v y) < e(z, v)$ and $e(z, w y) < e(z, w)$.
\end{proof}

Observe this also proves that we can restrict the proof to witnessed models, since for any modal formula in $\Gamma_P$, the value taken is no longer an infimum (respectively, supremum) but a minimum (maximum).

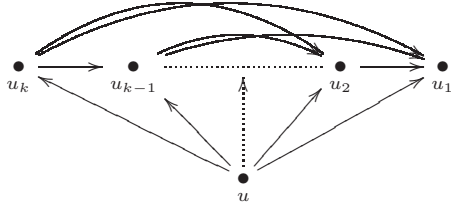
\begin{figure}
\vspace{1cm}
\begin{displaymath}
\scalebox{1}{
    \xymatrix{\underset{u_k}{\bullet} \ar[r] \ar@/^2pc/[rrr] \ar@/^2pc/[rrrr]& \underset{u_{k-1}}{\bullet} \ar@{.}[rr] \ar@/^1pc/[rr] \ar@/^1pc/[rrr] & &  \underset{u_2}{\bullet} \ar[r] & \underset{u_1}{\bullet}\\
    & & \underset{u}{\bullet} \ar[ull] \ar[ul] \ar[ur] \ar[urr] \ar@{.>}[u] & &}}
\end{displaymath}
\caption{Frame structure}\label{figureLocal}
\end{figure}

Our objective is now to prove completeness with respect to the class of linearly ordered models in the sense of Figure \ref{figureLocal}. Since from the previous lemma we get that the model is witnessed, intuitively we are only lacking to prove that, for a given world, we can select a particular unique successor (up to transitivity), and that this action preserves the value of the relevant formulas. Formula $(3)$ in $\Gamma_P$ takes care of this aspect.
\begin{lemma}\label{lemma:witness}
Let $\mod{M} \in 4\class{K}_{\aclass{A}}$ be finite tree with root $u$ such that $\Gamma_P \not \vdash_{\langle \mod{M}, u \rangle} \Box \varphi_P$, and let $z$ as in \ref{lemma:finiteDepthLocal}.
Then, for each $t \in W$ with $Rzt$ or $t=z$, and such that it has successors,  there is some world $t_{\mathrm{w}} \in W$ such that
$Rtt_{\mathrm{w}}$ and \[e(t, \Box v) = e(t_{\mathrm{w}}, v) \qquad \text{ and } \qquad e(t, \Box w) = e(t_{\mathrm{w}}, w).\]
\end{lemma} 
\begin{proof}
	
	Suppose towards a contradiction that there is not a common witness for $\Box v$ and $\Box w$, i.e., there are $r_1, r_2$ with $Rtr_1, Rtr_2$ and
	\begin{itemize}
		\item $e(t, \Box v) = e(r_1, v) = \alpha_y^{a_{r_1}}$,
		\item $e(t, \Box w) = e(r_2, w) = \alpha_y^{b_{r_2}}$,
		\item For any $r$ with $Rtr$, $a_r \leq a_{r_1}$ and $b_r \leq b_{r_2}$, and one of them is a strict inequality.
	\end{itemize}

	Then, for any $Rtr$, it holds that $e(r, v   w) \geq \alpha_y^{a_{r_1} + b_{r_2} -1}$, so $e(t, \Box(v   w)) \geq \alpha_y^{a_{r_1} + b_{r_2} -1}$. On the other hand, $e(t, \Box v   \Box w) = \alpha_y^{a_{r_1}}   \alpha_y^{b_{r_2}}$. 
	Now, for formula $(3)$ in $\Gamma_P$ to hold, it is necessary that $\alpha_y^{a_{r_1} + b_{r_2} -1} \leq \alpha_y^{a_{r_1} + b_{r_2}}$, and so, $\alpha_y^{a_{r_1} + b_{r_2} +n} = \alpha_y^{a_{r_1} + b_{r_2} -1}$ for any $n \in \omega$. However, this leads to have that $e(z, v   w) = \alpha_y^{a_{r_1} + b_{r_2} -1} =  \alpha_y^{a_{r_1} + b_{r_2} +n } = e(z, v   w)   \alpha_y$, which results in a contradiction since $e(z, \varphi_P) <1$. 
\end{proof}

Relying in the previous results, we can conclude a completeness lemma with respect to a very particular class of models: namely, with frames like in Figure \ref{figureLocal} and quite special evaluations.


%

Let us denote by 
$\what{4\class{K}_{\aclass{A}}}$ the class of models definable over frames with the structure in Fig. \ref{figureLocal}, i.e., for arbitrary but finite $n \in \omega$, 
\begin{itemize}
	\item $W = \{u_0, u_1, \ldots, u_n\}$ and
	\item $R = \{\langle u_i,u_j\rangle \colon$ for all $i \leq j\}$
\end{itemize}

Observe there is no bound on the size of the frames, while all of them are finite.

\begin{lemma}\label{lemma:completenessLocal}
	The following are equivalent:
	\begin{itemize}
		\item $\Gamma_P \vdash_{4\class{K}_{\aclass{A}}} \varphi_P$, 
		\item $\Gamma_P \vdash_{\scriptsize{\what{4\class{K}_{\aclass{A}}}}} \varphi_P$
	\end{itemize}
\end{lemma}
\begin{proof}
	Soundness is immediate. Concerning the left-to-right direction, assume there is a model $\mod{M} \in 4\class{K}_{\aclass{A}}$ with $u \in W$ be such that 
	$\Gamma_P \not \vdash_{\langle \mod{M}, u \rangle} \varphi_P$. 
	
Then consider the submodel $\what{\mod{M}}$ defined from $\mod{M}$ by taking its restriction to the universe  \[\what{W} = \bigcup_{i \in \N} \what{w_i}\] where
	\begin{itemize}
		\item $	\what{w_0} \coloneqq  \{u\}$
		\item 	$\what{w_1} \coloneqq  \{z\}$ as given in Lemma \ref{lemma:finiteDepthLocal}
		\item Let $\{t\} = \what{w_i}$. Then  put \[\what{w_{i+1}} \coloneqq  \begin{cases} \{t_{\mathrm{w}}\} \text{ as given in Lemma \ref{lemma:witness}} & \text{ if }t \text{ has any successors}\\  \emptyset &\text{ if }t \text{ has no successors}\end{cases}\]
	\end{itemize}
	It is a transitive model since the original $\mod{M}$ was so, and it clearly has the required frame (since $z$ had finite depth in the original model, for some $n$ onwards the set  $\what{w_n}$ will be empty).

	 Taking submodels does not change the value taken at each world by the propositional variables, 
	 i.e., for any $p \in \mathcal{V}$ (and thus, also for any non-modal formula) and any $t \in \what{W}$
	 it holds that $\what{e}(t,p) = e(t,p)$. Then we have that $\what{e}(z, \varphi_P) = e(z, \varphi_P) < 1$ (so $\what{e}(u, \Box \varphi_P) < 1$) and also that $\what{e}(u, \Box y) = \alpha_y = \what{e}(u, \Diamond y)$ (from Lemma \ref{lemma:valueylocal}), taking care of formula $(1)$ in $\Gamma_P$. 
	 
	 The remaining cases are the formulas with some modality and inside the scope of a $\Box$ operation in $\Gamma_P$, namely \begin{itemize}
	 	\item $\bigvee\limits_{1 \leq i \leq m} (v \leftrightarrow (\Box v)^{s^{\parallel \mathtt{v_i}\parallel}}  y^{\mathtt{v_i}}) \land  (w \leftrightarrow (\Box w)^{s^{\parallel \mathtt{w_i}\parallel}}  y^{\mathtt{w_i}})$ and
	 	\item $\Box (v    w) \to (\Box v    \Box w)$
	 \end{itemize}
	We just need to check that the values of those formulas are preserved from $\mod{M}$ to $\what{\mod{M}}$ in any world $t \in \what{W} \setminus \{u\}$. To do that, observe the only modal subformulas appearing are $\Box v$, $\Box w$ and $\Box (v   w)$, so it is enough to show the values of those three modal formulas are preserved.
	
	This can be easily done by induction in the depth (over the restricted model) of the world $t$.

	 \begin{itemize}
	 	\item If $\depth(t) = 0$, then also in $\mod{M}$ the world $t$ does not have successors, so clearly $1 = \what{e}(t, \Box \varphi) = e(t, \Box \varphi)$ for any formula $\varphi$.
	 	\item For $\depth(t) = n+1$, then also in $\mod{M}$ the world $t$ has successors, so $e(t, \Box v) = e(t_{\mathrm{w}}, v)$ from Lemma \ref{lemma:witness}, and we know that $\what{e}(t, \Box v) \leq \what{e}(t_{\mathrm{w}}, v) = e(t_{\mathrm{w}}, v)$ by Induction (since $t_{\mathrm{w}} \in \what{W}$. Moreover, it is clear that also $\what{e}(t, \Box v) \geq e(t, \Box v)$ given that $\what{\mod{M}}$ is a submodel of $\mod{M}$. Thus, $\what{e}(t, \Box v) = e(t, \Box v)$ and the same for what concerns $w$. Moreover, also $e(t, \Box (v \& w)) = e(t_{\mathrm{w}}, v \& w)$, so the same reasoning applies.
	 \end{itemize}

	\end{proof}

 It is an easy observation that whenever we use $(2)$ from $\Gamma_P$ to get that, at a certain point $r$ there is some $1 \leq i \leq m$ such that
 \[e(r, v) = e(r, \Box v)^{s^{\parallel \mathtt{v_i} \parallel}}   \alpha_y^{\mathtt{v_i}}\]
 there is in fact a unique such index $1 \leq i \leq m$, both for the value of $v$ and of $w$. Indeed, do not forget that $(2)$ determines with the same index the value of $v$ and that of $w$. Since there are no repetitions in $P$, for $1 \leq i \neq j \leq m$ it necessarily holds that either $\mathtt{v_i} \neq \mathtt{v_j}$ or $\mathtt{w_i} \neq \mathtt{w_j}$. Assuming any of those inequalities leads to have some $a \in \omega$ such that $\alpha_y^a$ is idempotent, and moreover, if we consider the inequality for the $\mathtt{v}$\scriptsize s \normalsize (and the same happens for $\mathtt{w}$), that $e(z, v) = \alpha_y^b$ for some $a \leq b$. Then, $e(z, v)   \alpha_y = e(z,v)$, contradicting once again $e(z,\varphi_P) <1$.

It is now natural to obtain an exact characterization of $v$ and $w$ in terms of 
$\alpha_y$ in each world of a model 
as in Figure \ref{figureLocal} satisfying $\Gamma_P$ in world $u_0$ and not satisfying $\varphi_P$ in that world . 
\begin{lemma}\label{lemma:charvwLocal}
Let $\mod{M} \in \what{4\class{K}_{\aclass{A}}}$ 
such that $\Gamma_P \not \vdash_{\langle \mod{M}, u\rangle} \varphi_P$ and $u_k$ is the element identified in Lemma \ref{lemma:finiteDepthLocal}. 

Then, for all $1 \leq j \leq k$
\[e(u_j, v) = \alpha_y^{\mathtt{v_{i_1}}\ldots \mathtt{v_{i_j}}} \qquad \text{ and } \qquad e(u_j, w) = \alpha_y ^{\mathtt{w_{i_1}}\ldots \mathtt{w_{i_j}}}\]
for $i_n$ being the unique value\footnote{This is properly defined, see the observation above.} in  $\{1, \ldots, m\}$ such that
\[e(u_n,v) = e(t, \Box v)^{s^{\parallel \mathtt{v_{i_n}}\parallel}}  \alpha_y^{\mathtt{v_{i_n}}} \quad \text{ and } \quad e(t,w) =  e(t, \Box w)^{s^{\parallel \mathtt{w_{i_n}}\parallel}}  \alpha_y^{\mathtt{w_{i_n}}}.\]

Moreover, for all $1 \leq j \leq k$,  \[e(u_j,v) = e(u_j,w) \text{ if and only if } \mathtt{v_{i_1}}\ldots \mathtt{v_{i_j}} = \mathtt{w_{i_1}}\ldots \mathtt{w_{i_j}}.\]
\end{lemma}
\begin{proof}

We will prove the first claim by induction on $j$. The details are only given for the $v$ case, the other one is proven in the same fashion.
\begin{itemize}
	\item If $j=1$ we know that $u_1$ has no successors, so from formula $(2)$ from $\Gamma_P$ we get
	\[e(u_1, v) = e(u_1, \Box v)^{s^{\parallel \mathtt{v_{i_1}} \parallel }}   e(u_1, y)^{\mathtt{v_{i_1}}} = \alpha_y^{\mathtt{v_{i_1}}}\]
	\item For $j = n+1$ using again formula $(2)$ we get that
	\[e(u_{n+1}, v) = e(u_{n+1}, \Box v)^{s^{\parallel \mathtt{v_{i_{n+1}}} \parallel }}   e(u_{n+1}, y)^{\mathtt{v_{i_{n+1}}}}\]
	
	From Lemma \ref{lemma:valuevpower} we get that $e(u_{n+1}, \Box v) = e(u_n, v)$ (observe the other worlds to which $u_{n+1}$ is related have all smaller depth, and so bigger values of $v$). Applying Induction Hypothesis we get the following chain of equalities
	\[e(u_{n+1}, v) = (\alpha_y^{\mathtt{v_{i_1}}\ldots \mathtt{v_{i_n}}})^{s^{\parallel \mathtt{v_{i_{n+1}}} \parallel }}   \alpha_y^{\mathtt{v_{i_{n+1}}}} =
	\alpha_y^{\mathtt{v_{i_1}}\ldots \mathtt{v_{i_n}} \cdot s^{\parallel \mathtt{v_{i_{n+1}}} \parallel }  + \mathtt{v_{i_{n+1}}}} =
	\alpha_y^{\mathtt{v_{i_1}}\ldots \mathtt{v_{i_n}}\mathtt{v_{i_{n+1}}}}.\]
\end{itemize}

Concerning the second claim, suppose towards a contradiction that there is $1 \leq j \leq k$ such that $\mathtt{v_{i_1}}\ldots \mathtt{v_{i_j}} \neq \mathtt{w_{i_1}}\ldots \mathtt{w_{i_j}}$ and 
$e(u_j, v) = \alpha_y^{\mathtt{v_{i_1}}\ldots \mathtt{v_{i_j}}} =  \alpha_y^{\mathtt{w_{i_1}}\ldots \mathtt{w_{i_j}}} =  e(u_j,w)$.
If $\mathtt{v_{i_1}}\ldots \mathtt{v_{i_j}} < \mathtt{w_{i_1}}\ldots \mathtt{w_{i_j}}$, from the residuated lattices properties it follows that
$\alpha_y^{\mathtt{v_{i_1}}\ldots \mathtt{v_{i_j}}}    \alpha_y = \alpha_y^{\mathtt{v_{i_1}}\ldots \mathtt{v_{i_j}}}$. 
It then follows that $e(u_j, v   y) = e(u_j, v)$ and trivially, that $e(u_k, v   y) = e(u_k, v)$. This contradicts 
$e(u_k, \varphi_P) < 1$, since this would require  that $e(u_k, v   y) < e(u_k, v)$

The analogous reasoning serves the case where $\mathtt{v_{i_1}}\ldots \mathtt{v_{i_j}} > \mathtt{w_{i_1}}\ldots \mathtt{w_{i_j}}$.
\end{proof}

It is now a simple observation that in a model as the one appearing in the above lemma, $e(u_k, \varphi_P) < 1$ implies that $e(u_k, v) = e(u_k, w)$, since either those two values are equal or there is some natural number $n > 1$ that
 $e(u_k, v) \leftrightarrow e(u_k, w) = \alpha_y^n \leq \alpha_y$, and so, making $e(u_k, \varphi_P) = 1$. 

%

%

Putting together all the previous results,  we can provide a completeness condition for the $\Gamma_P \vdash \varphi_P$ deductions.
\begin{corollary}\label{cor:completenessLocal}
	Assume $\Gamma_P \not \vdash_{4\class{K}_{\aclass{A}}} \varphi_P$. Then there is $\alg{A} \in \aclass{A}$, $k \in \omega$ and 
	$\mod{M} = \langle \{u, u_1, \ldots, u_k\}, \{\langle u, u_i \rangle:\? 1 \leq i \leq k\} \cup \{\langle u_i, u_j\rangle:\? 1 \leq j < i \leq k\}, e \rangle \in \what{\class{K}_{\{\alg{A}\}}}$ 
	such that 
	there exists a mapping $f \colon \{1\ldots k \} \to \{1 \ldots m\}$ and an element $\alpha \in A$ for which:
	\begin{itemize}
		\item For each $1 \leq j \leq k$, $e(u_j,v) = \alpha^{\mathtt{v}_{f(1)}\ldots \mathtt{v}_{f(j)}}$ 
		and $e(u_j,w) = \alpha^{\mathtt{w}_{f(1)}\ldots \mathtt{w}_{f(j)}}$
		\item $\mathtt{v}_{f(1)}\ldots \mathtt{v_{f(k)}} = \mathtt{w}_{f(1)}\ldots \mathtt{w}_{f(k)}$.
	\end{itemize}
\end{corollary}

It is now very natural to introduce the reduction itself from the Post Correspondence Problem to the local deduction over transitive models. Moreover, as we saw above, the reduction can be specified to finite models only.

\begin{proposition}\label{prop:PsatLocal}
	Let $P$ be an instance of the Post Correspondence Problem. Then the following are equivalent:
	\begin{enumerate}
		\item $P$ is satisfiable;
		\item $\Gamma_P \not \vdash_{4\class{K}_{\aclass{A}}} \Box \varphi_P$;
		\item $\Gamma_P \not \vdash_{\omega4\class{K}_{\aclass{A}}} \Box \varphi_P$.
	\end{enumerate}
\end{proposition}
\begin{proof}
	Trivially $(3)$ implies $(2)$. Moreover, Lemma \ref{lemma:completenessLocal} proves that $(2)$ implies $(3)$.

	On the other hand, the fact that $(3)$ implies $(1)$  follows immediately from Corollary \ref{cor:completenessLocal}. Indeed, if $\Gamma_P \not \vdash_{4\class{K}_{\class{A}}} \varphi_P$, then from that corollary we know there is some $k$ and map $f \colon \{1\ldots k \} \to \{1 \ldots m\}$ such that 
	$f(1), \ldots, f(k)$ is a solution for $P$.

	To prove that $(1)$ implies $(3)$ assume that $P$ has a solution $i_1, \ldots, i_k$, and assume without loss of generality that there is no $j < k$ such that $i_1, \ldots, i_j$ is a solution too. 
	By assumption, there is some $\alg{A} \in \class{A}$ such that $\alg{A}$ is not $2 \cdot (\mathtt{v}_{i_1}\ldots\mathtt{v}_{i_k})$-contractive,
	so there is some element $\alpha$ for which 
	$\alpha^{2 \cdot (\mathtt{v}_{i_1}\ldots\mathtt{v}_{i_k}) +1}< \alpha^{2 \cdot (\mathtt{v}_{i_1}\ldots\mathtt{v}_{i_k})}$. 
	
	Then define the Kripke model $\mod{M} = \langle W, R, e \rangle$ by letting 
	\begin{itemize}
		\item $W = \{u, u_1, \ldots, u_k\}$,
		\item $R = \{\langle u, u_i \rangle \colon 1 \leq i \leq k\} \cup \{\langle u_i, u_j \rangle \colon 1 \leq j < i \leq k \}$,
		\item For each $1 \leq j \leq k$, define the evaluation at each $u_j$, for $1 \leq j \leq k$, by:\footnote{ The evaluation of variables in $u$ is irrelevant to the evaluation of $\Gamma_P, \varphi_P$.}
		\begin{itemize}
			\item $e(u_j, y) = \alpha$, 
			\item $e(u_j, v ) = \alpha^{\mathtt{v_{i_1}}\ldots \mathtt{v_{i_j}}}$, 
			\item $e(u_j, w) = \alpha^{\mathtt{w_{i_1}}\ldots \mathtt{w_{i_j}}}$
		\end{itemize}
	\end{itemize}

	It is now a matter of simple calculations to see that $\mod{M}$ globally validates the formulas from $\Gamma_P$. On the other hand, observe that  $e(u_k, v \leftrightarrow w) = 1$ ($i_1, \ldots, i_k$ was a solution for P). Since $e(u_k, y) = \alpha < 1$, and 
	$e(u_k, (v   w \rightarrow v   w   y) < 1$ for all $1 \leq j \leq k$ (since $\alpha$ was chosen non $2 \cdot (\mathtt{v}_{i_1}\ldots\mathtt{v}_{i_k})$-contractive), this gives us that $e(u_k, \varphi_P) < 1$, concluding the proof.
\end{proof}

Theorem \ref{th:undec} results as a direct corollary of the previous result.

\section{Modal \L ukasiewicz logics}\label{sec:lukDec}
We can now turn our attention to two of the modal fuzzy logics studied in the previous section: the ones arising respectively from $[0,1]_{\sL}$ and from $\{MV_n\colon n \in \omega\}$. We will see in this way some interesting phenomena that are revealed when comparing the minimal modal logics and their corresponding transitive versions.

Interestingly enough, we can prove that the logic $\vdash_{K_\ssL}$ is decidable. 
To the best of our knowledge, examples of logics turning undecidable when transitivity is involved affect more complex situations, referring for instance to the addition of a transitive closure operator to predicate logics \cite{Gan99}, \cite{GraOtRo99}, or related to very expressive logics that include forward and backward accessibility relations and also allow a certain level of quantification \cite{Zo17}. The case of study here shows a relatively surprising example of a decidable local deduction whose transitive extension is undecidable. 

In order to prove decidability of $\vdash_{K_\ssL}$ it is crucial the continuity of all underlying propositional operations, which will leads to a good behaviour of the \L ukasiewicz Kripke models. It can be proven that the logic $\vdash_{\class{K}_\ssL}$ is complete with respect to witnessed models, by relying in the analogous result for predicate (standard) \L ukasiewicz logics (\cite{Ha05b},\cite{Ca18}).

To prove the completeness of the modal logic wrt witnessed models, it is only necessary to use the natural translation from modal into predicate logics and back. Since it is lacking in the literature, we proceed with the details, but the main technical issue is the analogous proof of completeness in first order standard \L ukasiewicz. No previous knowledge on the topic is required to proceed, through some observations and results from \cite{Ha98} \cite{Ha05b} and \cite{Ca18} will be used. 
 
Recall that, given a type of relations $\{R_i\}$ of respective arity $ar(R_i)$, a (standard FO) \L ukasiewicz model  is a structure
  \[\mod{M} = \langle D_{\mod{M}}, \{R_i^{\mod{M}}\}_{i \in I}\rangle\] where $R_i^{\mod{M}} \colon D_{\mod{M}}^{ar(R_i)} \mapsto [0,1]$. For a certain
  		 formula $\varphi(\overline{x})$ we write $\varphi[\overline{a}]^{\mod{M}}$ to denote the value taken by $\varphi$ in the structure under any  evaluation that sends $\overline{x}$ to $\overline{a}$, defined inductively by 

		\begin{itemize}
			\item $R_i[\overline{a}]^{\mod{M}} = R_i^{\mod{M}}(\overline{a})$, 
			\item $(\psi \divideontimes \chi)[\overline{a}]^{\mod{M}} = \psi[\overline{a}]^{\mod{M}} \divideontimes \chi[\overline{a}]^{\mod{M}}$ for $\divideontimes$ propositional connective, 
			\item $(\exists x \varphi)[\overline{a}]^{\mod{M}} = \sup_{w \in W} \varphi[w,\overline{a}]^{\mod{M}}$,
			\item $(\forall x \varphi)[\overline{a}]^{\mod{M}} = \inf_{w \in W} \varphi[w,\overline{a}]^{\mod{M}}$.
		\end{itemize} 
Since the \L ukasiewicz negation is involutive, we have that $\forall x \varphi(x) = \neg \exists x \neg \varphi(x)$ (and correspondingly, in the modal logic, $\Box \varphi = \neg \Diamond \neg \varphi$), so we will be referring below only to the existential quantifier (and respectively, to the $\Diamond$ modal operator).

A \termDef{$\L \forall$-embedding} of an structure $\mod{M}$ into a structure $\mod{N}$ is a mapping $h \colon D_{\mod{M}} \rightarrow D_{\mod{N}}$ such that for any first order formula $\varphi$, and any $\overline{a} \in D_{\mod{M}}^{ar(\varphi)}$ it holds that \[\varphi[\overline{a}]^{\mod{M}} = \varphi[h(\overline{a})]^{\mod{N}}.\] In particular, the valuation of sentences is preserved.

Moreover, we say that a structure $\mod{M}$ is \termDef{witnessed} (analogously to the definition  for Kripke models) whenever for any formula $\exists x \varphi(x, \overline{y})$ and any $\overline{a}$ tuple of $\vert \overline{y} \vert$ elements of $D_{\mod{M}}$ there is some $b \in D_{\mod{M}}$ for which
\[(\exists x \varphi)[\overline{a}]^\mod{M} = \varphi[b,\overline{a}]^\mod{M}.\]

On the other hand, given two $[0,1]_\sL$ Kripke models $\mod{M}$ and $\mod{N}$, a mapping $h\colon W_{\mod{M}} \rightarrow W_{\mod{N}}$ is a \termDef{$\L K$-embedding} whenever for any modal formula $\varphi$ and any $v \in W_{\mod{M}}$, it holds that \[e(v, \varphi) = e(h(v), \varphi).\]

\begin{lemma}
	\emph{(\cite[Lemma~3]{Ha05b} and \cite[Prop.~3.10]{Ca18})}
	Any standard FO \L ukasiewic model can be $\L \forall$-embedded in a witnessed one. 
\end{lemma}

From that, we can easily get the analogous result for $K_\sL$.
\begin{lemma}
Any \L ukasiewic Kripke model $\mod{M}$ can be $\L K$-embedded in a witnessed one. 
\end{lemma}
\begin{proof}
	For a (countable) set of variables $\mathcal{V}$, let $\tau$ be the predicate language $\{R/2, \{P/1\colon p \in \mathcal{V}\}\}$. It s clear that we can stablish a bijection between \L ukasiewicz Kripke models ($\mod{M}$) and (standard FO) \L ukasiewicz models ($\mod{M}'$) on language $\tau$ that satisfy the sentence $C_R\colon \forall x, y (R(x,y) \vee \neg R(x,y))$, in such a way that the two structures share domain and for each $\varphi$ modal formula and each $v \in W$ it holds
	\[e(v, \varphi) = \varphi^\sharp[v]^{\mod{M}'}\]
	where  $\psi^\sharp(x)$ is the standard translation into FO, i.e., 
	\begin{eqnarray*}
	p^\sharp(x) &\coloneqq& P(x)\\
	 (\varphi \star \psi)^\sharp(x) &\coloneqq& \varphi^\sharp(x) \star \psi^\sharp(x) \text{ for }\star \in \{ , \rightarrow, \wedge,...\}\\
	(\Diamond \varphi)^\sharp(x) &\coloneqq& \exists y (R(x,y)   \varphi^\sharp(y)).
	\end{eqnarray*}

	Simply take the same domain, and let \[Rvw \Longleftrightarrow R[v,w]^{\mod{M}'} = 1 \quad \text{ and } \quad e(v,p) = P[v]^{\mod{M}'}.\]

Then, consider a \L ukasiewicz Kripke models $\mod{M}$, and its corresponding FO model $\mod{M}'$. The previous lemma gives us a witnessed FO model $\mod{N}'$ in which $\mod{M}$ can be $\L \forall$-embedded with a mapping $\sigma$. In particular, true sentences are preserved, so $\mod{N}' \models C_R$. Thus, we can use the previous bijection and refer to the Kripke model $\mod{N}$ over domain $D_{\mod{N}'}$ and such that $e_{\mod{N}}(v, \varphi) = \varphi^\sharp[v]^{\mod{N}'}$ for any modal formula $\varphi$. 

It is easy to see also that $\mod{N}$ is witnessed too. For pick a modal formula $\Diamond \varphi$ and a world $v$ in the universe of $\mod{N}$. If $e(v, \Diamond \varphi) = 0$ it  is trivially witnessed (by any related world). Otherwise, we have the following chain of equalities:
\begin{eqnarray*}
e_{\mod{N}}(v, \Diamond \varphi) = (\Diamond \varphi)^\sharp [v]^{\mod{N}'} =
 (\exists y R[v,y]   )\varphi^\sharp(y))^{\mod{N}'} &\overset{\mod{N}' \text{witnessed}}{=}&\\
  (R[v,w]   \varphi^\sharp[w])^{\mod{N}'} = e_{\mod{N}}(w, \varphi) \text{ and }Rvw \text{ (in }\mod{N})
\end{eqnarray*}

Clearly, the same mapping $\sigma$ that was a $\L \forall$-embedding from $\mod{M}'$ to $\mod{N}'$ is also a $\L K$-embedding from $\mod{M}$ to $\mod{N}$, since for any formula $\varphi$ and any $v \in W_{\mod{M}}$, 
\begin{eqnarray*}
e_{\mod{M}}(v, \varphi) = \varphi^{\sharp}[v]^{\mod{M}'} = \varphi^{\sharp}[\sigma(h)]^{\mod{N}'} = e_{\mod{N}}(\sigma(v), \varphi).
\end{eqnarray*}
\end{proof}

\begin{corollary}\label{cor:witcomplete}
	$\vdash_{K_\ssL}$ is complete with respect to witnessed models.
\end{corollary}

From here, it is not hard to prove decidability of $\vdash_{K_\ssL}$, in a similar fashion to the procedure given in \cite[Def.~3]{Ha05b}. 

Fix $\Gamma \cup \{\varphi\} \subset_{\omega} Fm$. Let 
\begin{itemize}
	\item $\Sigma_0 \coloneqq \{\Diamond\chi \in \PSF(\Gamma \cup \{\varphi\})$, 
	\item $\Sigma_{i+1} \coloneqq \{\Diamond \chi \in \PSF(\{\psi\colon \Diamond \psi \in \Sigma_i\})\}$.
\end{itemize}
Observe that, since $\Gamma \cup\{\varphi\}$ is a finite set it has a maximum modal depth degree $N$, and $\Sigma_i$ is empty for all $i \geq N$.

Let then $W_0 \coloneqq \{w_{\langle 0 \rangle}\}$, $W_{i+1} \coloneqq \{w_{\langle \sigma, \Diamond \chi \rangle}\colon w_{\langle \sigma\rangle} \in W_i, \Diamond \chi \in \Sigma_i\}$, and let $W \coloneqq \bigcup_{i < N} W_i$. Observe $W$ is a finite set. Our goal is to use $w_{\langle \sigma, \Diamond \chi \rangle}$ for witnessing the value of $\Diamond \chi$ at world $w_{\langle \sigma \rangle}$.

Assume $\mathcal{V}$ is the set of propositional variables of $\Gamma \cup \{\varphi\}$. Let us define $\mathcal{V}^\Diamond$ as the following extended set of propositional variables combining the two previous notions and the original set $\mathcal{V}$:
\begin{itemize}
	\item $x_{w}$ for each $x \in \mathcal{V}$, $w \in W$, 
	\item $\overline{\Diamond \psi}_{w}$ for each $\Diamond \psi  \in \Sigma_i$ and $w \in W_i$, for $1 \leq i < N$.
\end{itemize}

We will now use the previous language to define a set of propositional formulas that will determine intrinsically the same conditions that hold in a corresponding Kripke model. To do that, 
let us first define a translation from the original modal formulas (in $\mathcal{V}$) to the natural correspondent over $\mathcal{V}^\Diamond$. 

Let \begin{itemize}
	\item $\const{0}^\sharp(w) \coloneqq \const{0}$, $\const{1}^\sharp(w) \coloneqq \const{1}$,
	\item $x^\sharp(w) \coloneqq x_w$ for $x \in \mathcal{V}$,
	\item $(\psi \star \chi)^\sharp(w) \coloneqq \psi^\sharp(w) \star \chi^\sharp(w)$ for $\star$ propositional connective ($ , \rightarrow$)
	\item $(\Diamond \psi)^\sharp(w) \coloneqq \overline{\Diamond \psi}_w$
\end{itemize}

Observe that, by construction, the set $\Gamma^\sharp(w_0) \cup \{\varphi^\sharp(w_0)\}$ is a finite set of propositional \L ukasiewicz formulas in the set of variables $\mathcal{V}^\Diamond$.

Let us now define the set of formulas $\Psi(\Gamma \cup \{\varphi\})$ that will determine the behaviour of modal formulas/variables, as the union of
\[\overline{\Diamond \psi}_{w_{\langle \sigma \rangle}} \leftrightarrow \psi^\sharp(w_{\langle \sigma, \Diamond \psi\rangle }) \quad \text{ and }\quad \bigvee_{w_{\langle \sigma, \Diamond \chi\rangle}\in W} \psi^\sharp(w_{\langle \sigma, \Diamond \chi \rangle}) \rightarrow \psi^\sharp(w_{\langle \sigma, \Diamond \psi\rangle })\]

for each $\overline{\Diamond \psi}_{w_{\langle \sigma \rangle }} \in \mathcal{V}^\Diamond$. Observe that if $\overline{\Diamond \psi}_{w_{\langle \sigma \rangle }} \in \mathcal{V}^\Diamond$, then for any  $w_{\langle \sigma, \Diamond \chi\rangle}\in W$, the formula $ \psi^\sharp(w_{\langle \sigma, \Diamond \chi \rangle})$ is in the language of $\mathcal{V}^\Diamond$. Thus $\Psi(\Gamma \cup \{\varphi\})$ is also a finite set of propositional \L ukasiewicz formulas in the set of variables $\mathcal{V}^\Diamond$.

\begin{lemma}\label{lem:KlukToLuk}
	$\Gamma \vdash_{K_\ssL} \varphi$ if and only if $\Gamma^{\sharp}(w_0), \Psi(\Gamma \cup \{\varphi\}) \vdash_{\sL} \varphi^\sharp(w_0)$.\footnote{Where $\vdash_{\sL}$ denotes the usual propositional (finitary) \L ukasiewicz $[0,1]$-valued logic}
\end{lemma}
\begin{proof}
	To prove left to right direction assume $\Gamma \not \vdash_{K_\ssL} \varphi$. From Corollary \ref{cor:witcomplete} we know there is a witnessed model $\mod{M}$ and $w \in W$ such that $e(w, \Gamma) = 1$ and $e(w, \varphi) < 1$. Since it is witnessed, for each formula $\Diamond \psi$ and each world $v \in W$ it holds that there is some world $v_{\Diamond \psi}$
	such that $Rvv_{\Diamond \psi}(v, \Diamond\psi)$ and $e(v, \Diamond \psi) = e(v_{\Diamond \psi}, \psi)$. Let us denote $w$ by $w_{\langle 0\rangle}$ and, inductively from $w_{\langle 0\rangle}$, let $w_{\langle \sigma, \Diamond \psi\rangle}$ denote the world ${w_{\langle \sigma \rangle}}_{\Diamond \psi}$.

	Then, consider the mapping $h \colon \mathcal{V}^\Diamond \rightarrow [0,1]$ given by 
\begin{itemize}
	\item $h(x_{w_{\langle \sigma \rangle}}) = e(w_{\langle \sigma \rangle}, x)$ for $x \in \mathcal{V}$, 
	\item $h(\overline{\Diamond \psi}_{w_{\langle \sigma \rangle}}) = e(w_{\langle \sigma \rangle}, \Diamond \psi)$.
\end{itemize}	
It is clear that $h(\Gamma^\sharp(w_0)) = e(w_0, \Gamma) = 1$ and $h(\varphi^\sharp(w_0)) = w(w_0, \varphi) <1$. On the other hand, since the model is witnessed \[h(\overline{\Diamond \psi}_{w_{\langle \sigma \rangle}}) = e(w_{\langle \sigma \rangle}, \Diamond \psi) = e(w_{\langle \sigma, \Diamond \psi\rangle}, \psi)\] and so, $h(\Psi(\Gamma \cup \{\varphi\})) = 1$ too, proving that $\Gamma^{\sharp}(w_0), \Psi(\Gamma \cup \{\varphi\}) \not \vdash_{\sL} \varphi^\sharp(w_0)$.

For what concerns left to right direction, the construction of the Kripke model from a propositional homomorphism $h \colon \mathcal{V}^\Diamond \rightarrow [0,1]_{\sL}$ that sends the premises to $1$ and the conclusion to less than $1$ is immediate. Simply, define the universe of the model by $W$ as introduced above, and let $R = \{\langle w_{\langle \sigma\rangle}, w_{\langle \sigma, \Diamond \psi\rangle} \rangle \colon w_{\langle \sigma, \Diamond \psi\rangle} \in W\}$. Moreover, let $e(w_{\langle \sigma \rangle}, x) = h(x_{w_{\langle \sigma \rangle}})$ for each $x \in \mathcal{V}$. It is simple to prove by induction that for any modal formula $\psi$ and any $w \in W$ such that $\psi^\sharp(w)$ is in variables $\mathcal{V}^\Diamond$, it holds that 
\[e(w, \psi) = h(\psi^\sharp(w))\]
It is trivial for the propositional connectives. For what concerns the modal formulas, observe that by definition
\[e(w_{\langle \sigma \rangle}, \Diamond \psi) = \bigvee_{Rw_{\langle \sigma \rangle}v} e(v, \psi) =  \bigvee_{w_{\langle \sigma, \Diamond \chi \rangle} \in W} e(w_{\langle \sigma, \Diamond \chi \rangle}, \psi).\]
By Induction Hypothesis, this equals to $ \bigvee_{w_{\langle \sigma, \Diamond \chi \rangle} \in W} h(\psi^\sharp_{w_{\langle \sigma, \Diamond \chi \rangle}})$. But from the formulas in $\Psi(\Gamma \cup \{\varphi\})$ we know that $h(\psi^\sharp_{w_{\langle \sigma, \Diamond \chi \rangle}}) \leq h(\psi^\sharp_{w_{\langle \sigma, \Diamond \psi \rangle}})$ for all such worlds, so in particular we get that
\[e(w_{\langle \sigma \rangle}, \Diamond \psi) = h(\psi^\sharp_{w_{\langle \sigma, \Diamond \psi \rangle}})\]
From $\Psi(\Gamma \cup \{\varphi\})$ it also holds that $h(\overline{\Diamond \psi}_{w_{\langle \sigma \rangle}}) = h(\psi^\sharp_{w_{\langle \sigma, \Diamond \psi \rangle}})$, concluding the proof.

	\end{proof}

Since it is well known that $\vdash_{\sL}$ is decidable \cite{Ha98}, the following is immediate.
\begin{corollary}
	The finitary companion of $\vdash_{K_\ssL}$ is decidable.
\end{corollary}

A second observation concerns the relation between the modal logics arising from the standard MV algebra ($\vdash_{\class{K}_{\ssL}}$) and from the family of all finite MV algebras ($\vdash_{\class{K}_{\omega\ssL}}$). 
It is well known that at a propositional level, the two logics coincide (see eg. \cite{Ha98}). This fact, in combination with Lemma \ref{lem:KlukToLuk} above, give us a direct proof of the fact that the (minimal) local modal logic arising from $\class{K}_{\sL}$ and the one arising from $\class{K}_{\{MV_n\colon n \in \omega\}}$ coincide too. Indeed, while it is immediate that $\vdash_{\class{K}_{\ssL}} \subseteq \vdash_{4\class{K}_{\omega\ssL}}$, the other inclusion comes using the same construction of a Kripke model from a propositional homomorphism  that sends the premises to $1$ and the conclusion to less than $1$, simply taking now $h \colon \mathcal{V}^\Diamond \rightarrow MV_n$ for some suitable (big enough) $n$.

Surprisingly enough, the corresponding transitive logics do not coincide, as the following construction shows.
	
	\begin{lemma} The following hold:
	\begin{itemize}
		\item $x \leftrightarrow (\Box x)^2, \Box(x \leftrightarrow (\Box x)^2), \neg \Diamond \Box \perp \vdash_{4\class{K}_{\omega\ssL}} \neg x \vee x $ and 
		\item $x \leftrightarrow (\Box x)^2, \Box(x \leftrightarrow (\Box x)^2), \neg \Diamond \Box \perp \not \vdash_{4\class{K}_{\ssL}} \neg x \vee x$.
	\end{itemize}	
%
\end{lemma}
\begin{proof}
On the one hand, it is not hard to find a model validating the second statement. Indeed, let $\mod{M} \coloneqq \langle \omega, \{\langle n,m\rangle \colon n < m \in \omega\}, e\rangle$ with $e(0, x) = 0.1$ (any arbitrary value in $(0,1)$ serves our porpoise) and
\[e(n+1, x) = \frac{e(n, x) +1}{2}\]
Clearly, $e(n, \Box \perp) = 0$ for all $n$, since each world has a successor, and trivially $e(0,x \vee \neg x) < 1$. 

For each world in the model, it is easy  to see that $e(n,x) < 1$ for all $n \in \omega$, and that $n < m$ implies $e(n,x) < e(m,x)$, so $e(n,\Box x) = e(n+1,x)$ for all $n \in \omega$. In particular, $e(n, (\Box x)^2) = 2 \frac{e(n, x) +1}{2} -1 = e(n,x)$, proving that $e(0,x \leftrightarrow (\Box x)^2) = 1 = e(0,\Box (x \leftrightarrow (\Box x)^2))$.

On the other hand, suppose there is $n \in \omega$ and $\mod{M}$ a transitive model over $MV_n = \{0, \frac{1}{n}, \ldots, \frac{n}{n}\}$, with $v$ a world of the model in which $e(v, x) = \frac{l}{n}$ with $0 < l < n$, $l \in \omega$. Assume further that $e(v, \neg \Diamond \Box \perp) = 1$, so any successor of $v$ has also some successor world. For the other premise to hold in $v$, there must be some sequence of worlds $\{v_i\colon i \in \omega\}$ with $v_0 = v$, $Rv_iv_{i+1}$ and $e(v_i, x) = e(v_{i+1},x^2)$. However, for this sequence it would then hold $e(v_i, x) < e(v_{i+1}, x)$, while having $e(v_i, x) < 1$ for all $i$ (otherwise, the whole sequence would evaluate $x$ to $1$ and so would do the initial world $v$). Since $MV_n$ has finitely  many elements, this increasing sequence cannot exist, proving our claim.
	
\end{proof}

It can be proven that the previous example also serves to differentiate $\vdash_{4\class{K}_{\Pi}}$ and the transitive modal logic over a one-generated subalgebra of $[0,1]_\Pi$. However, we do not know whether their corresponding minimal modal logics (not transitive) coincide.

\begin{corollary}
$\vdash_{\class{K}_{\omega\ssL}}$ coincides with $\vdash_{\class{K}_{\ssL}}$, while $\vdash_{4\class{K}_{\omega\ssL}}$ is strictly stronger than $\vdash_{4\class{K}_{\ssL}}$. 
\end{corollary}

A consequence of this fact is that it cannot exist a set of axioms and rules $G4$ such that both 
\begin{itemize}
	\item the extension of $\vdash_{\class{K}_{\ssL}}$ with $G4$ coincides with $\vdash_{4\class{K}_{\ssL}}$, and
	\item the extension of $\vdash_{\class{K}_{\omega\ssL}}$ with $G4$ coincides with $\vdash_{4\class{K}_{\omega\ssL}}$.
\end{itemize}

In particular, usual axiom $4: \Box \varphi \rightarrow \Box \Box \varphi$ is no longer enough to characterize transitive models of the class in at least one of the previous cases.

\section{The presence of $\Delta$}\label{sec:delta}

As in fragments of predicate logics (see eg. \cite{BaCiFe07}), in the presence of the projection operation $\Delta$ we can translate the undecidability results to the set of theorems of the respective logics, and also to the local SAT problem\footnote{Given a formula, is there some model  and some world in it that evaluates the formula to $1$?} -since, with $\Delta$, the problems of validity and local SAT are easily reducible one to the other, contrary to the situation without $\Delta$.

The observation is totally natural, but nevertheless, relevant for what concerns possible applications of these logics, since in practical uses, the possibility  of talk about absolute truth of a formula seems reasonable. However, the fact that in its presence we can more easily fall in undecidable questions gives an idea of the possible step in expressibility power taken when adding $\Delta$ to the language.

Monteiro-Baaz $\Delta$ operation is defined, for an arbitrary $FL_{ew}$-chain by letting \[\Delta(a) = \begin{cases} 1&\hbox{ if } a = 1\\ 0  &\hbox{ otherwise.}\end{cases}\]
Then, the Deduction Theorem, not necessarily holding in the modal logics studied in Section 3\footnote{Observe not even the usual local DT (analogous to the one holding in propositional $\Pi$ and $\L$ logics) seems natural to prove: while for each particular model it is true that $\gamma \models_{\mod{M}} \varphi$ iff there is some $n \in \omega$ such that $\models_{\mod{M}} \gamma^n \rightarrow \varphi$, this index may vary from one model to the other, and in particular, the family might fail to have a supremum in $\omega$. A deeper study of this question is left for future works.} is fully recovered. Indeed, we have that for any class $\class{C}$ of models evaluated over $FL_{ew}$-chains, 
\[\gamma \vdash_{\class{C}} \varphi \text{ if and only if } \vdash_{\class{C}} \Delta \gamma \rightarrow \varphi\]

Allow us to write $\vdash_{\class{C}}^{\Delta}$ to denote the logic over the class of models $\class{C}$ whose language has been expanded by the $\Delta$ operation interpreted (at each world) as described above.

\begin{lemma}
	\begin{enumerate}
		\item The set of valid formulas of $\vdash_{4\class{K}_{\aclass{A}}}^{\Delta}$ is undecidable. Moreover, the set of valid formulas of $\vdash_{\omega 4\class{K}_{\aclass{A}}}^{\Delta}$ is also undecidable.
	\item The problems of local SAT in $4\class{K}_{\aclass{A}}$ and in $\omega 4\class{K}_{\aclass{A}}$ with $\Delta$ are undecidable. 
\end{enumerate}
\end{lemma}
\begin{proof}
$\mathit{(1)}$ follows naturally from the DT and Theorem \ref{th:undec}. 
	For the second, it is trivial that $\varphi$ is valid in $\vdash_{4\class{K}_{\aclass{A}}}^{\Delta}$ (resp. $\omega \vdash_{4\class{K}_{\aclass{A}}}^{\Delta}$) if and only if $\neg \Delta \varphi$ is not locally SAT in $4\class{K}_{\aclass{A}}$ (resp. $\omega 4\class{K}_{\aclass{A}}$) with $\Delta$. 
	\end{proof}

\section{Conclusions and Future work}
We have studied the computability of a large family of transitive modal many-valued logics, proving their undecidability. Moreover, we have compared the behaviour of the transitive \L ukasiewicz modal logics (over $[0,1]_\sL$ and over $\{MV_n \colon n \in \omega\}$) and their corresponding transitive versions, observing some particular behaviours that contrast with the known results in other modal logics. 

Several interesting open problems are remaining after this study. First natural question is whether transitive modal Gödel logic (over models with a crisp accessibility, in particular) is decidable, which would provide a full understanding of the three main left-continuous t-norm based logics. In ongoing works we are studying this question, non trivial from \cite{CaMeRo17} since the logic is not necessarily  complete with respect to models of finite depth.

On the other hand, the question of whether the local modal product logic with crisp-accessibility models is decidable or not also remains open. In particular, the proof from \cite{CeEsBo10} concerning decidability of SAT and theoremhood questions over the analogous logic over valued-accessibility models seems hardly adaptable to the crisp case, since it is crucial in the proof to allow the accessibility relation to be valued in $(0,1)$.

\section{Acknowledgements}

This project has received funding from the European Union's Horizon 2020 research and innovation program under the Marie Sklodowska-Curie grant agreement No 689176 (SYSMICS project) and by the grant no. CZ.02.2.69/0.0/0.0/17\_050/0008361 of the Operational programme Research, Development, Education of the Ministry of Education, Youth and Sport of the Czech Republic, co-financed by the European Union.

\bibliographystyle{abbrv}

\begin{thebibliography}{10}
	
	\bibitem{BaPe11a}
	F.~Baader and R.~Pe{\~n}aloza.
	\newblock {GCIs} make reasoning in fuzzy {DL} with the product t-norm
	undecidable.
	\newblock In {\em Proceedings of the 2011 International Workshop on Description
		Logics, DL'11,}, 2011.
	
	\bibitem{BaCiFe07}
	M.~Baaz, A.~Ciabattoni, and C.~G. Ferm{\"u}ller.
	\newblock Monadic fragments of {G\"o}del logics: Decidability and
	undecidability results.
	\newblock In {\em LPAR, Logic for Programming, Artificial Intelligence, and
		Reasoning, 14th International Conference, LPAR 2007, Yerevan, Armenia,
		October 15-19, 2007, Proceedings}, volume 4790 of {\em Lecture Notes in
		Computer Science}, pages 77--91, 2007.
	
	\bibitem{BoDiPe15}
	S.~Borgwardt, F.~Distel, and R.~Pe{\~n}aloza.
	\newblock The limits of decidability in fuzzy description logics with general
	concept inclusions.
	\newblock {\em Artificial Intelligence}, 218:23--55, 2015.
	
	\bibitem{BoEsGoRo11}
	F.~Bou, F.~Esteva, L.~Godo, and R.~Rodr{\'\i}guez.
	\newblock On the minimum many-valued modal logic over a finite residuated
	lattice.
	\newblock {\em Journal of Logic and Computation}, 21(5):739--790, 2011.
	
	\bibitem{Ca18}
	X.~Caicedo.
	\newblock Lindstr{\"o}m theorems for {{\L}}ukasiewicz predicate logic.
	\newblock {\em Fundamenta Mathematicae}, (To appear).
	
	\bibitem{CaMe13}
	X.~Caicedo, G.~Metcalfe, R.~Rodr{\'\i}guez, and J.~Rogger.
	\newblock A finite model property for {G}\"odel modal logics.
	\newblock In L.~Libkin, U.~Kohlenbach, and R.~de~Queiroz, editors, {\em Logic,
		Language, Information, and Computation}, volume 8071 of {\em Lecture Notes in
		Computer Science}. Springer Berlin Heidelberg, 2013.
	
	\bibitem{CaMeRo17}
	X.~Caicedo, G.~Metcalfe, R.~Rodr{\'\i}guez, and J.~Rogger.
	\newblock Decidability of order-based modal logics.
	\newblock {\em Journal of Computer and System Sciences}, 88:53 -- 74, 2017.
	
	\bibitem{CaRo10}
	X.~Caicedo and R.~O. Rodr{\'\i}guez.
	\newblock Standard {G}{\"o}del modal logics.
	\newblock {\em Studia Logica}, 94(2):189--214, 2010.
	
	\bibitem{CaRo15}
	X.~Caicedo and R.~O. Rodriguez.
	\newblock Bi-modal {G}{\"o}del logic over $[0,1]$-valued {K}ripke frames.
	\newblock {\em Journal of Logic and Computation}, 25(1):37--55, 2015.
	
	\bibitem{CeEsBo10}
	M.~Cerami, F.~Esteva, and F.~Bou.
	\newblock Decidability of a description logic over infinite-valued product
	logic.
	\newblock In F.~Lin, U.~Sattler, and M.~Truszczynski, editors, {\em Principles
		of Knowledge Representation and Reasoning: Proceedings of the Twelfth
		International Conference, {KR} 2010, Toronto, Ontario, Canada, May 9-13,
		2010}, pages 203--213. AAAI Press, 2010.
	
	\bibitem{CeEs18}
	M.~Cerami, F.~Esteva, and A.~Garcia-Cerda\~na.
	\newblock On the relationship between fuzzy description logics and many-valued
	modal logics.
	\newblock {\em International Journal of Approximate Reasoning}, 93:372--394,
	2018.
	
	\bibitem{CeSt13}
	M.~Cerami and U.~Straccia.
	\newblock On the undecidability of fuzzy description logics with {GCI}'s with
	{{\L}}ukasiewicz t-norm.
	\newblock {\em Information Sciences}, 227:1--21, 2013.
	
	\bibitem{DoSH93}
	K.~Do{\v s}en and P.~Schroeder-Heister, editors.
	\newblock {\em Substructural Logics}, volume~2 of {\em Studies in Logic and
		Computation}.
	\newblock Oxford University Press, 1993.
	
	\bibitem{EsGo01}
	F.~Esteva and L.~Godo.
	\newblock Monoidal t-norm based logic: towards a logic for left-continuous
	t-norms.
	\newblock {\em Fuzzy Sets and Systems}, 124:271--288, 2001.
	
	\bibitem{Fi92a}
	M.~Fitting.
	\newblock Many-valued modal logics.
	\newblock {\em Fundamenta Informaticae}, 15:235--254, 1992.
	
	\bibitem{Fi92b}
	M.~Fitting.
	\newblock Many-valued modal logics, {II}.
	\newblock {\em Fundamenta Informaticae}, 17:55--73, 1992.
	
	\bibitem{GaJiKoOn07}
	N.~Galatos, P.~Jipsen, T.~Kowalski, and H.~Ono.
	\newblock {\em Residuated Lattices: an algebraic glimpse at substructural
		logics}, volume 151 of {\em Studies in Logic and the Foundations of
		Mathematics}.
	\newblock Elsevier, Amsterdam, 2007.
	
	\bibitem{Gan99}
	H.~Ganzinger, C.~Meyer, and M.~Veanes.
	\newblock The two-variable guarded fragment with transitive relations.
	\newblock In {\em Proceedings. 14th Symposium on Logic in Computer Science
		(Cat. No. PR00158)}, pages 24--34, 1999.
	
	\bibitem{GraOtRo99}
	E.~Gr{\"a}del, M.~Otto, and E.~Rosen.
	\newblock {Undecidability Results on Two-Variable Logics}.
	\newblock {\em Archive for Mathematical Logic}, 38:213--354, 1999.
	
	\bibitem{Ha98}
	P.~H{\'a}jek.
	\newblock {\em Metamathematics of fuzzy logic}, volume~4 of {\em Trends in
		Logic---Studia Logica Library}.
	\newblock Kluwer Academic Publishers, Dordrecht, 1998.
	
	\bibitem{Ha05b}
	P.~H{\'a}jek.
	\newblock Making fuzzy description logic more general.
	\newblock {\em Fuzzy Sets and Systems}, 154(1):1--15, 2005.
	
	\bibitem{HaTe13}
	G.~Hansoul and B.~Teheux.
	\newblock Extending {\l}ukasiewicz logics with a modality: Algebraic approach
	to relational semantics.
	\newblock {\em Studia Logica}, 101(3):505--545, 2013.
	
	\bibitem{MeOl11}
	G.~Metcalfe and N.~Olivetti.
	\newblock Towards a proof theory of {G}{\"o}del modal logics.
	\newblock {\em Logical Methods in Computer Science}, 7(2):27, 2011.
	
	\bibitem{Post46}
	E.~L. Post.
	\newblock A variant of a recursively unsolvable problem.
	\newblock {\em Bulletin of the American Mathematical Society}, pages 264--268,
	1946.
	
	\bibitem{St01}
	U.~Straccia.
	\newblock Reasoning within fuzzy description logics.
	\newblock {\em Journal of Artificial Intelligence Research}, 14:137--166, 2001.
	
	\bibitem{ViEsGo16}
	A.~Vidal, F.~Esteva, and L.~Godo.
	\newblock On modal extensions of product fuzzy logic.
	\newblock {\em Journal of Logic and Computation}, 27(1):299--336, 2017.
	
	\bibitem{Zo17}
	E.~Zolin.
	\newblock Undecidability of the transitive graded modal logic with converse.
	\newblock {\em Journal of Logic and Computation}, 27(5):1399--1420, 2017.
	
\end{thebibliography}

\end{document}